\documentclass[twocolumn,aps,superscriptaddress,nofootinbib,floatfix,linenumbers]{revtex4}
\usepackage{epsfig,bm,feynmf}
\usepackage{graphics}
\usepackage{amsmath}
\usepackage{mathrsfs}
\usepackage{appendix}
\usepackage{bm}
\usepackage[normalem]{ulem}  % \sout{old text} for strikeout
\usepackage[dvips]{color} % For blue in-text comments and additions

\renewcommand{\sout}{\bgroup \color{red} \ULdepth=-.5ex \ULset}

\begin{document}
\title{Probing the topological charge in QCD matter via multiplicity up-down asymmetry}

% authors
\author{Yifeng Sun} 
%\email{sunyfphy@physics.tamu.edu}
\email{sun112358@comp.tamu.edu}
\affiliation{Cyclotron Institute and Department of Physics and Astronomy, Texas A$\&$M University, College Station, Texas 77843, USA}%

\author{Che Ming Ko}
\email{ko@comp.tamu.edu}
\affiliation{Cyclotron Institute and Department of Physics and Astronomy, Texas A$\&$M University, College Station, Texas 77843, USA}%

% date
\date{\today}% It is always \today, today,

\begin{abstract}
Relativistic heavy ion collisions provide the possibility to study the topological charge in QCD matter through the event-by-event fluctuating net axial charge or nonequal numbers of left- and right-handed quarks they generate in the produced quark-gluon plasma.  Based on the chiral kinetic approach for nearly massless quarks and antiquarks in the strong vorticity field produced along the normal direction of the reaction plane of non-central heavy ion collisions, we show that a unique signal for the topological charge in QCD matter can be identified from the asymmetric distribution of particles with momenta pointing in the upper and lower hemispheres of the reaction plane as a result of the fluctuating net axial charge.
\end{abstract}
%\keywords{Chiral vortical effect,  chiral kinetic approach, relativistic heavy ion collisions, AMPT}

\maketitle

\section{Introduction}

The topological charge in QCD matter~\cite{PhysRevD.30.2212,PhysRevD.36.581,Moore2011,PhysRevD.93.074036} can be studied in relativistic heavy ion collisions through the effect of the event-by-event fluctuating net axial charge or non-equal numbers of left- and right-handed quarks they generate in the produced quark-gluon plasma~\cite{Kharzeev:2007jp}.  In the presence of the magnetic field created in non-central heavy ion collisions, the finite net axial charge can lead to a separation of positively and negatively charged particles in the transverse plane of a collision as a result of the vector charge current it generates along the direction of the magnetic field~\cite{Kharzeev:2007jp}.  This so-called chiral magnetic effect (CME)~\cite{Kharzeev:2007jp,PhysRevD.78.074033,KHARZEEV2010205}, which has been observed in condensed matter systems such as the Weyl semimetals in external magnetic fields~\cite{Li:2014bha} and studied in other areas of physics~\cite{Kharzeev:2013ffa}, has been suggested as a possible explanation for the observed charge separation in experiments~\cite{PhysRevLett.103.251601,PhysRevC.81.054908,PhysRevLett.110.012301,PhysRevLett.118.122301}.  Realistic studies of this effect based on anomalous hydrodynamics~\cite{YIN201642,Hirono:2014oda,Jiang:2016wve,Shi:2017cpu} and chiral kinetic approach~\cite{HUANG2018177,Sun:2018idn} require a magnetic field of lifetime of at least $\tau_{B}=$ 0.6 fm$/c$ to account for the experimental data. Such a long-lived magnetic field does not seem to be supported by the small electric conductivity of QGP from lattice QCD calculations~\cite{PhysRevD.83.034504,PhysRevLett.99.022002}.

On the other hand,  the vorticity field produced in non-central heavy ion collisions also affects quarks and antiquarks of right-handness differently from those of left-handness, although independent of their charges, and it decays slowly with time~\cite{PhysRevC.94.044910}. Similar to the separation of charges due to the CME, the vorticity field can lead to a separation of baryons and antibaryons with respect to the reaction plane of a heavy ion collision~\cite{PhysRevLett.106.062301,PhysRevD.92.071501}. This chiral vortical effect (CVE) depends, however, also on the net baryon in the produced QGP, which is unfortunately very small in relativistic heavy ion collisions and thus makes the signal of CVE hard to detect. Since  the vorticity field not only leads to a partial alignment of the spins of both positively and negatively charged quarks and antiquarks along its direction, as evidenced by the observed spin polarization of $\Lambda$ hyperons in these collisions~\cite{STAR:2017ckg}, but also tends to make quarks and antiquarks of opposite handedness to move in opposite directions.  With different numbers of right- and left-handed quarks and antiquarks in the quark matter, there will appear a difference between the numbers of quarks and antiquarks that move along and opposite to the direction of the vorticity field, resulting in a multiplicity up-down asymmetry with respect to the reaction plane of a heavy ion collision.  

In the present paper, this effect is studied quantitatively in the chiral transport approach with initial quark and antiquark distributions taken from a multiphase transport model~\cite{PhysRevC.72.064901}, which includes the essential collision dynamics of relativistic heavy ion collisions through its fluctuating initial conditions and strong partonic scatterings. Our results show that the multiplicity up-down asymmetry induced by the strong vorticity field created in the direction normal to the reaction plane of a heavy ion collision depends sensitively on the net axial charge fluctuation in the produced quark matter, providing thus a more promising probe to the topological charge in QCD matter than the CME and CVE.  
 
This paper is organized as follows. In Sec. II, we briefly review the equations of motion of nearly massless quarks and anitquarks and their scatterings in the chiral kinetic approach. The initial conditions and the magnetic and vorticity fields that are needed for carrying out the chiral kinetic calculations are described in Sec. III. Results on the multiplicity up-down asymmetry and its event-by-event distribution in Au + Au collisions at $\sqrt{s_{NN}}=62.4$ GeV and centrality of 30-40\% are given in Sec. IV to illustrate the effect of the vorticity field on the luctuating net axial charge in the produced partonic matter as a result of its nonzero topological charge.  Finally, a summary is given in Sec. V.
 
\section{The chiral kinetic approach}

In the chiral transport approach to massless quarks and antiquarks in both magnetic $\bf B$ and vorticity $\boldsymbol\omega$ fields, their equations of motion are given by~\cite{PhysRevLett.109.162001,PhysRevD.96.016002,PhysRevC.94.045204,PhysRevC.96.024906}
\begin{eqnarray}
&&\dot{\mathbf{r}}=\frac{\hat{\mathbf{p}}+Q\lambda(\hat{\mathbf{p}}\cdot\mathbf{b})\mathbf{B}+2\lambda p(\hat{\mathbf{p}}\cdot\mathbf{b})\boldsymbol{\omega}}{1+Q\lambda\mathbf{b}\cdot\mathbf{B}+6\lambda p(\mathbf{b}\cdot\boldsymbol{\omega})},\label{CKM}\\
&&\dot{\mathbf{p}}=\frac{Q\hat{\mathbf{p}}\times\mathbf{B}}{1+Q\lambda\mathbf{b}\cdot\mathbf{B}+6\lambda p(\mathbf{b}\cdot\boldsymbol{\omega}),}\label{LF}
\end{eqnarray}
where $Q$ and $\lambda=\pm 1$ are the charge and helicity of a quark or antiquark (parton), and $\mathbf{b}=\frac{\mathbf{p}}{2p^3}$ is the Berry curvature that results from the adiabatic approximation of taking the spin of a massless parton to be always parallel or anti-parallel to its momentum. Corrections to above equations due to the small light $u$ and $d$ quark maasses ($m_u=3$ MeV and $m_d=$ 6 MeV)~\cite{Olive:2016xmw} can be included by replacing $\hat{\mathbf{p}}$, $p$ and $\mathbf{b}$ with $\frac{\mathbf{p}}{E_p}$, $E_p$ and $\frac{\hat{\bf{p}}}{2E_p^2}$, respectively, as in Ref.~\cite{PhysRevD.89.094003}.

The factor $\sqrt{G}=1+Q\lambda\mathbf{b}\cdot\mathbf{B}+6\lambda p(\mathbf{b}\cdot\boldsymbol{\omega})$ in the denominator of Eqs.(\ref{CKM}) and (\ref{LF}) modifies the phase-space distribution of partons and ensures the conservation of vector charge.  The modified parton equilibrium distribution can be achieved from parton scatterings by 
requiring the parton momenta $\mathbf{p}_3$ and $\mathbf{p}_4$ after a two-body scattering, which is determined by their total scattering cross section, with the probability $\sqrt{G(\mathbf{p}_3)}\sqrt{G(\mathbf{p}_4)}$~\cite{PhysRevC.96.024906}.  For the parton scattering cross section $\sigma_{\rm tot}$, we choose it to reproduce the small shear viscosity to entropy density ratio $\eta/s$ in QGP extracted from  experimentally measured anisotropic flows in relativistic heavy ion collisions based on the viscous hydrodynamics~\cite{PhysRevLett.99.172301,PhysRevC.78.024902} and transport models~\cite{FERINI2009325,PhysRevC.79.014904}. This empirically determined value is close to the conjectured lower bound for a strongly coupled system in conformal field theory~\cite{PhysRevLett.94.111601} and the values from lattice QCD calculations~\cite{Nakamura:2004sy}.  For a partonic matter dominated by light quarks as considered here, we can relate $\eta/s$ to the total cross section $\sigma_{\rm tot}$ by $\eta/s=\frac{1}{15}\langle p\rangle \tau=\frac{\langle p\rangle}{10n\sigma_{\rm tot}}$~\cite{PhysRevD.90.114009} if the cross section is taken to be isotropic, where $\tau$ is the relaxation time of the partonic matter, $n$ is the parton number density, and $\langle p\rangle$ is the average momentum of partons. Taking $\eta/s=1.5/4\pi$ as determined in Ref.~\cite{PhysRevC.85.024901} from anisotropic flows in relativistic heavy ion collisions using viscous hydrodynamics, we then calculate the parton scattering cross section as a function of parton density and temperature or energy density. 

\section{Initial conditions and the magnetic and vorticity fields}

For the initial phase-space distribution of partons, we take it from the string melting version of the AMPT model~\cite{PhysRevC.72.064901} with the values $a=0.5$ and $b=0.9$ GeV$^2$ in the Lund string fragmentation function to give a better description of the charged particle multiplicity density, momentum spectrum, and two- and three-particle correlations~\cite{PhysRevC.84.014903,SUN2017219} in heavy ion collisions at RHIC.   For the event-by-event fluctuating net axial charges in heavy ion collisions due to the topological charge fluctuation in QCD~\cite{PhysRevD.30.2212,PhysRevD.36.581,Moore2011,PhysRevD.93.074036}, we let each event to have either more right-handed quarks and antiquark or left-handed quarks and antiquarks with the probability $(1+p)/2$, where  
$p=\sqrt{\langle N_5^2\rangle}/N$ with $\sqrt{\langle N_5^2\rangle}$ being the  initial axial charge fluctuation and $N$ is the total number of partons in an event. 

For the magnetic field, we obtain it from the Lienard-Wiechert potential produced by the spectator protons in the colliding nuclei. As shown in Refs.~\cite{BZDAK2012171,PhysRevC.85.044907}, the resulting magnetic field in the overlap region of the two nuclei is in the direction perpendicular to the reaction plane and has a very large strength but a very short lifetime.  Since  the partonic effect on the magnetic field is small~\cite{PhysRevC.97.044906}, due to the small electric conductivity in QGP from lattice QCD calculations~\cite{PhysRevD.83.034504,PhysRevLett.99.022002}, we neglect it in the present study and also assume that the magnetic field is uniform in space.  As to the vorticity field, it is calculated from the velocity field $\mathbf{v}({\bf r},t)$ of partons, which is determined from the average velocity of partons in a local cell of the partonic matter  via $\boldsymbol{\omega}=\frac{1}{2}\boldsymbol{\nabla}\times\mathbf{u}$ with $\mathbf{u}=\gamma\mathbf{v}$ and $\gamma=\frac{1}{\sqrt{1-\mathbf{v}^2}}$ as described in Ref.~\cite{PhysRevC.96.024906}. We note that contrary to the short lifetime of the magnetic field, the vorticity field produced in non-central heavy ion collisions decays slowly with time~\cite{PhysRevC.94.044910}. 

The partonic matter from the AMPT model after including event-by-event fluctuations in the net axial charge is then evolved according to the chiral kinetic equations of motion and parton scatterings described in the above until its energy density decreases to $\epsilon_0=0.56$ GeV$/$fm$^3$, similar to the critical energy density from LQCD for the partonic to hadronic transition~\cite{BORSANYI201499} and also that corresponding to the switching temperature $T_{\rm SW}=$ 165 MeV from the partonic to the hadronic phase used in viscous hydrodynamics~\cite{PhysRevC.83.054912}.

\section{Results}

To illustrate the effect of vorticity field, we consider Au$+$Au collisions at $\sqrt{s_{NN}}=62.4$ GeV and centrality of 30-40$\%$ for the two cases of without ($p=0$) and with ($p=0.4$) initial axial charge fluctuation in the partonic matter.  For the latter, only events of more right- than left-handed partons are considered.

In the presence of both magnetic and vorticity fields and with a net axial charge density in a partonic matter, the distributions of the azimuthal angles ($\phi$) of positively and negatively charged partons can be expressed by
\begin{eqnarray} 
\frac{dN_{\pm}}{d\phi}\propto 1&+&2v_2\cos(2\phi-2\Psi_{\rm RP}) \nonumber
\\&+&2(a_{\rm{CVE}}\pm a_{\rm{CME}})\sin(\phi-\Psi_{\rm RP}),
\label{aa}
\end{eqnarray}
where  $\Psi_{\rm RP}$ is the azimuthal angle of the reaction plane in a collision, $v_2$ is the elliptic flow, and $a_{\rm{CVE}}$ and $a_{\rm{CME}}$ are the multiplicity up-down asymmetry of charged partons induced by the vorticity and magnetic fields, respectively. Both $a_{\rm{CVE}}$ and $a_{\rm{CME}}$ can be positive or negative depending on the sign of the net axial charge  in the partonic matter.  Because of the opposite effects of magnetic field on the multiplicity up-down asymmetry of positively and negatively charged partons, we can study the effect of vorticity field by considering the azimuthal angle distribution of both positively and negatively charged quarks and antiquarks to remove the contribution of $a_{\rm CME}$ in Eq.(\ref{aa}).

\subsection{The multiplicity up-down asymmetry}

\begin{figure}[h]
\centering
\includegraphics[width=0.9\linewidth] {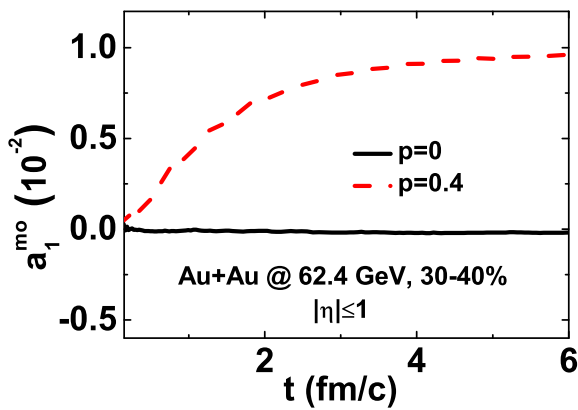}
\caption{(Color online) Time evolution of the multiplicity up-down asymmetry $a_1^{\rm mo}$ of mid-pseudorapidity ($|\eta|\le1$) light quarks in momentum space in  Au$+$Au collisions at $\sqrt{s_{NN}}=62.4$ GeV and centrality of 30-40$\%$ for the two cases of without ($p=0$) and with ($p=0.4$) initial net axial charge fluctuation. }
\label{a1p}
\end{figure}

Figure~\ref{a1p} shows the time evolution of the multiplicity up-down asymmetry $a_1^{\rm mo}=\langle \sin\phi\rangle=a_{\rm{CVE}}$ of mid-pseudorapidity ($|\eta|\le$1) light quarks and antiquarks in momentum space, where the average is over all events.  The black line is the result for the case of $p=0$, i.e., zero net axial charge fluctuation $\sqrt{\langle N_5^2\rangle}/N=0$.  In this case, momenta of light quarks in mid-pseudorapidity have no preferential directions in the transverse plane of a collision. For the case of $p=0.4$, shown by the red line, the finite net axial charge fluctuation leads to an increase of the multiplicity up-down asymmetry $a_1^{\rm mo}$ with time. 

\begin{figure}[h]
\centering
\includegraphics[width=0.9\linewidth] {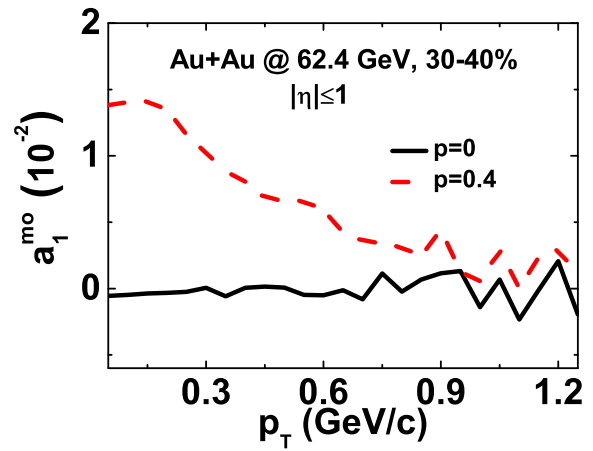}
\caption{(Color online) Same as Fig.~\ref{a1p} for the transverse momentum dependence of the multiplicity up-down asymmetry $a_1^{\rm mo}$ of light quarks.}
\label{pt}
\end{figure}

Results for the transverse momentum dependence of the multiplicity up-down asymmetry $a_1^{\rm mo}$ of light quarks in momentum space are shown in Fig.~\ref{pt}, again from events with more right- than left-handed quarks. It is seen that the $a_1^{\rm mo}$ for the case of $p=0$ is consistent with zero for all transverse momenta.  With non-zero net axial charge  fluctuation of $p=0.4$, $a_1^{\rm mo}$ becomes non-zero and decreases with transverse momentum. The latter is due to the decrease of the Berry curvature $\mathbf{b}=\frac{\mathbf{p}}{2p^3}$ in Eqs.(\ref{CKM}) and (\ref{LF}) with increasing parton transverse momentum.

\subsection{Multiplicity up-down asymmetry event distribution}

\begin{figure}[h]
\centering
\includegraphics[width=0.9\linewidth] {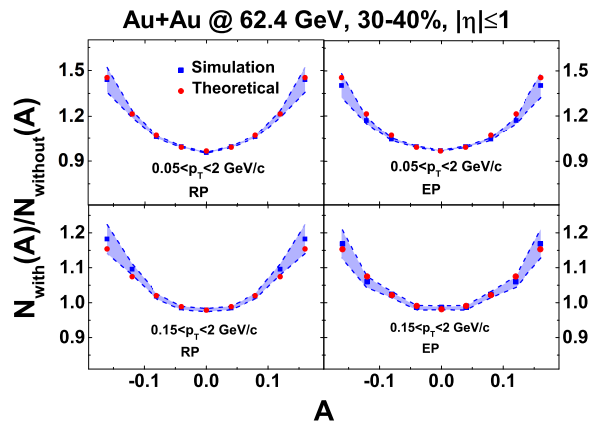}
\caption{(Color online) Ratio of multiplicity up-down asymmetry event distributions for events with [$N_{\rm with}(A)$] and without [$N_{\rm without}(A)$] net axial charge fluctuation for different particle transverse momentum ranges in Au+Au collisions at $\sqrt{s_{NN}}=62.4$ GeV and centrality 30-40$\%$. Red solid circles are obtained assuming a normal distribution for the multiplicity up-down asymmetry $A$.}
\label{ratio}
\end{figure}

Although the multiplicity up-down asymmetry $a_1^{\rm mo}$ is nonzero and positive for events with more right- than left-handed quarks and antiquarks obtained from a finite initial axial charge fluctuation $p$, it is negative with same magnitude if events with more left- than right-handed quarks and antiquarks are obtained with the same value of $p$. As a result, its average value over all events $\langle a_1^{\rm mo}\rangle$ is zero as the average of net axial charge $N_5$ is zero.    On the other hand, the event-by-event distribution $N(A)$ of the normalized multiplicity up-down asymmetry $A=\frac{N_U-N_D}{N_U+N_D}$, where $N_U$ and $N_D$ are the numbers of partons with momenta pointing in the upper and lower hemispheres of the reaction plane, respectively, is wider as $\langle a_{\rm{CVE}}^2\rangle$ becomes larger.  We therefore introduce the multiplicity up-down asymmetry event distribution $N(A)$ and consider the ratio $N_{\rm with}(A)/N_{\rm without}(A)$ of those with [$N_{\rm with}(A)$] and without [$N_{\rm without}(A)$] a net axial charge fluctuation.  In upper panels of Fig.~\ref{ratio}, we show by the blue band this ratio for mid-pseudorapidity light quarks of transverse momenta in the range of $0.05<p_T<2$ GeV$/c$. The upper left panel is for the case using initial reaction plane of a collision, and it shows a distinct concave shape.  Using the event plane determined by the azimuthal angles of emitted particles leads to the same conclusion as shown by the blue band in the upper right panel of Fig.~\ref{ratio}. For particles in the smaller transverse momentum range of $0.15<p_T<2$ GeV$/c$, their multiplicity up-down asymmetry event distribution ratio $N_{\rm with}(A)/N_{\rm without}(A)$, shown in lower panels of Fig.~\ref{ratio}, also has a concave shape whether the initial reaction plane or the event plane is used, although its curvature is smaller than the for partons in the larger transverse momentum range. 

The above results can be understood as follows. According to the particle azimuthal angle distribution given in Eq.~(\ref{aa}), a particle has the probabilities $\frac{1+a}{2}$ and $\frac{1-a}{2}$ to have a positive and negative value for $\rm{sin}(\phi-\Psi_{RP})$, repectively, where $a=4a_{\rm{CVE}}/\pi$.  The event-by-event distribution of the up-down asymmetry $A$ then has zero average value and a fluctuation given by 
\begin{eqnarray} 
\langle(\Delta A)^2\rangle=\frac{1}{N}+\frac{N-1}{N}a^2, 
\end{eqnarray}
where $N=N_U+N_D$. Taking into account the fluctuation of particle number $N$ in each event for a given momentum range, the fluctuation of $A$ 
is thus 
\begin{eqnarray} 
\langle\langle(\Delta A)^2\rangle\rangle=\left\langle\frac{1}{N}\right\rangle+\left(1-\left\langle\frac{1}{N}\right\rangle\right)a^2. 
\end{eqnarray}

If the number of particles $N$ is sufficient large and the particles in each event have no correlations, the final distribution of $A$ then has the normal Gaussian distribution $\mathcal{N}(0,\langle\langle(\Delta A)^2\rangle\rangle)$ according to the central limit theorem.  From the value $\langle{1/N}\rangle=2.133\times 10^{-3}$ and $\langle a_{\rm{CVE}}\rangle=\pm 9.82\times 10^{-3}$ for mid-pseudorapidity light quarks in the transverse momentum range of $0.05<p_T<2$ GeV$/c$ for the collisions considered in present study, where the plus and minus signs are for events with more right-handed particles and more left-handed quarks, respectively, and the average is taken over particles in corresponding events, the resulting ratio $N_{\rm with}(A)/N_{\rm without}(A)$ is shown by red solid circles in upper left panel of Fig.~\ref{ratio}.  They are seen to be similar to those calculated with $p=0.4$ and $p=0$ for mid-pseudorapidity light quarks in the same transverse momentum range within the uncertainty of using the normal distribution for the multiplicity up-down asymmetry $A$. This similarity is also seen in the case of using the event plane from final particles, as shown in the upper right panel of Fig.~\ref{ratio}, as well as for particles with momenta in the range of $0.15<p_T<2$ GeV$/c$, where $\langle{1/N}\rangle=2.914\times 10^{-3}$ and $\langle a_{\rm{CVE}}\rangle=\pm 8.29\times 10^{-3}$, as shown in lower panels of Fig.~\ref{ratio}.

\begin{figure}[h]
\centering
\includegraphics[width=0.9\linewidth] {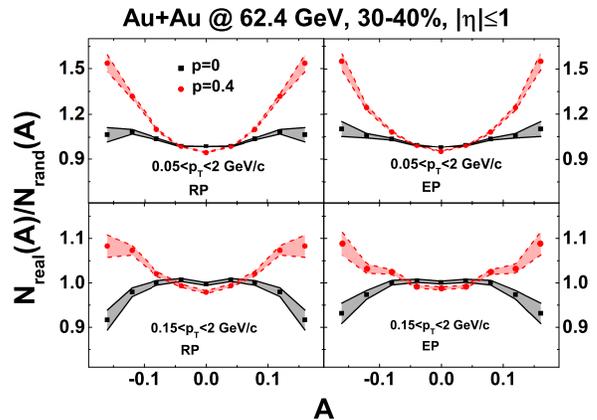}
\caption{(Color online) Ratio of multiplicity up-down asymmetry event distribution to $N_{\rm real}(A)$  to that with random up-down momentum distributions $N_{\rm rand}(A)$ for mid-pseudorapidity light quarks in different transverse momentum ranges in Au+Au collisions at $\sqrt{s_{NN}}=62.4$ GeV and centrality of 30-40$\%$.}
\label{ra}
\end{figure}

To help identify this effect in experiments, we follow the consideration of Ref.~\cite{PhysRevC.97.061901} on charge separation due to the CME by introducing the ratio $N_{\rm{real}}(A)/N_{\rm{rand}}(A)$, where $N_{\rm{real}}(A)$ is the real distribution of $A$ and $N_{\rm{rand}}(A)$ is the distribution of $A$ with the momentum of each parton in an event having the same probability to be in the upper and lower hemispheres of the reaction plane. In upper panels of Fig.~\ref{ra}, we show this ratio for mid-pseudorapidity light quarks of transverse momenta in the range of $0.05<p_T<2$ GeV$/c$. The upper left panel is for the case using initial reaction plane of a collision, and it shows a more distinct concave shape for this ratio for the case with a nonzero axial charge fluctuation than the case with zero axial charge fluctuation except the statistical fluctuation. Using the event plane determined by the azimuthal angles of emitted particles leads to the same conclusion as shown in the upper right panel of Fig.~\ref{ra}. For particles in the smaller transverse momentum range of $0.15<p_T<2$ GeV$/c$, their $N_{\rm{real}}(A)/N_{\rm{rand}}(A)$ ratio, shown in lower panels of Fig.~\ref{ra}, for the case without net axial charge fluctuation shows a convex shape whether the initial reaction plane or final event plane is used.  Including a nonzero axial charge fluctuation in the partonic matter, this ratio changes to the concave shape. We note that the convex shape for the $N_{\rm{real}}(A)/N_{\rm{rand}}(A)$ ratio in the case without net axial charge fluctuation is due to the fact that partons in the string melting version of the AMPT model are from decays of hadrons produced in the HIJING model~\cite{PhysRevD.44.3501}, which is used as its initial conditions, and are thus partially correlated in momentum space even after undergoing multiple scatterings. Without any momentum correlations, the $N_{\rm{real}}(A)/N_{\rm{rand}}(A)$ ratio should have a constant value of one. 

\section{Summary}

To summarize, we have proposed to study the effect of the vorticity field  in non-central relativistic heavy ion collisions on the particle multiplicity up-down asymmetry relative to the reaction plane in order to probe the net axial charge fluctuation in the produced partonic matter, which is related to the topological charge in QCD.  Solving the chiral transport equation in the presence of a self-consistent voriticity field and including the vector charge conserved scatterings among quarks and antiquarks from the AMPT model, we have found that the multiplicity up-down asymmetry, which is quantified by the multiplicity up-down asymmetry event distribution $N(A)$, which is related to the distribution of the difference between the numbers of quarks and antiquarks with momenta pointing in the upper and lower hemispheres of the reaction plane, is sensitive to the net axial charge fluctuation in the partonic matter.  In particular, the ratio between the multiplicity up-down asymmetry event distributions for the cases with finite and zero net axial charge fluctuation is directly related to the inverse of the multiplicity of partons and the net axial charge fluctuation, besides depending on the strength of the vorticity field.  Because of its local structure~\cite{Becattini:2017gcx,Xia:2018tes}, the vorticity field has been shown to result in a local spin polarization of $\Lambda$ hyperon in the direction of total orbital angular momentum that can be as large as 10\%~\cite{Xia:2018tes} and also depends less on the collision energy.  This is in contrast to the global spin polarization, which has a magnitude of about 3\% at $\sqrt{s_{NN}}=$7.7 GeV and decreases with the collision energy.  Since the multiplicity up-down asymmetry distribution depends on the square of the strength of vorticity field, its sensitivity to the net axial charge can be significantly enhanced if, for example, particles with momentum components in the event plane satisfying $p_xp_z>0$ or $p_xp_z<0$ are used in the analysis.  Measuring the multiplicity up-down asymmetry event distribution in non-central relativistic heavy ion collisions thus provides a promising method to probe the net axial charge fluctuation in partonic matter and thus the topological charge in QCD matter.  

\section*{Acknowledgements}

This work was supported in part by the US Department of Energy under Contract No. DE-SC0015266 and the Welch Foundation under Grant No. A-1358.

\bibliography{ref}

\end{document}